\documentclass[traditabstract,preprint]{aa}

\usepackage[utf8]{inputenc}
\usepackage{graphicx}
\usepackage{amsmath}
\usepackage{comment}
\usepackage{multirow}
\usepackage[varg]{txfonts}
\usepackage{subcaption}

\usepackage{natbib}
\bibpunct{(}{)}{;}{a}{}{,}

\providecommand{\abs}[1]{\lvert#1\rvert}

\begin{document}

\title{Robust analysis of differential Faraday Rotation based on interferometric closure observables.}

\author{Albentosa-Ruiz, E.
\and Marti-Vidal, I.}
\institute{Dpt. Astronomia i Astrof\'isica, Universitat de Val\`encia, C/ Dr. Moliner 50, 46120 Burjassot (Spain) \and Observatori Astron\`omic, Universitat de Val\`encia, C/ Cat. Jos\'e Beltr\'an 2, 46980 Paterna (Spain).}

\date{\today}

\abstract{Polarization calibration of interferometric observations is a costly procedure and, in some cases (e.g., a limited coverage of parallactic angle for the calibrator), it may not be possible to be performed. To avoid this worst-case scenario and expand the possibilities for the exploitation of polarization interferometric observations, the use of a new set of calibration-independent quantities (the closure traces) has been proposed. However, these quantities suffer from some degeneracies, so their use in practical situations may be rather limited. In this paper, we explore the use of closure traces on simulated and real observations, and show that (with the proper selection of fitting parameters) it is possible to retrieve information of the source polarization using only closure traces and constrain spatially resolved polarization. We carry out the first application of closure traces to the brightness modelling of real data, using the ALMA observations of M87 conducted on the April 2017 EHT campaign, quantifying a gradient in the Faraday rotation (FR) along the source structure (the M87 jet). This work opens the possibility to apply similar strategies to observations from any kind of interferometer (with a special focus on VLBI), from which quantities like differential Rotation Measure (RM) or the spatially resolved polarization can be retrieved.}


\titlerunning{Differential Faraday rotation with closure traces}
\authorrunning{Albentosa-Ruiz \& Marti-Vidal}

\maketitle

\section{Introduction}

Interferometric techniques \citep[e.g.][]{thompson2017interferometry} have proven to be essential for obtaining high-resolution images of celestial radio sources, since the long baselines used enable the synthesis of large apertures. However, to properly analyse the interferometric data, a thoughtful calibration and reduction is required, as both atmospheric effects and instrumental limitations modulate the phase and amplitude of the observed signals, affecting the interferometric observations and the images that are generated from them.

These atmospheric and instrumental effects are especially relevant at high observing frequencies in VLBI, particularly in polarimetric observations, since the resolutions are so high that it is very difficult to find point-like calibrators. The effect of the structure further complicates the estimation of the instrumental polarization, which is an important limiting factor that increases the difficulty of the calibration at mm/sub-mm wavelengths \citep[see e.g.][]{EHTVII}. Even though the calibration of these instrumental effects is often difficult, there are observable quantities (the so-called {\em closures}) that encode information about the properties of the observed sources while being rather insensitive to most of the atmospheric and instrumental effects. Closure quantities have been used in the analysis of interferometric observations since almost the beginning of phase-sensitive long-baseline astronomical radio interferometry \citep[e.g.][]{Pearson1984} and most of the image reconstruction algorithms rely on them, either indirectly \citep[via the ``self-calibration'' used as part of the hybrid imaging techniques][]{ReadheadAndWilkinson1978} or directly \citep[as part of the error function used in direct deconvolution algorithms, e.g.][]{Akiyamaetal2017SIMILIa,Akiyamaetal2017SIMILIb}. Recently, a new set of complex closure quantities, dubbed \textit{closure traces}, has been reported \citep[see][]{Broderick&Pesce2020}. The closure traces are complex closure quantities constructed from the visibility matrices $\pmb{V}$ measured for baselines connecting four antennas $\{A,B,C,D\}$ (see more details, along with an analytical study of the closure traces, in Ap. \ref{Ap:Section_1.5}), as follows:
\begin{equation}
\label{closure_trace}
\mathcal{T}_{ABCD}=\frac{1}{2} tr(\pmb{V}_{AB}\pmb{V}^{-1}_{CB}\pmb{V}_{CD}\pmb{V}^{-1}_{AD}).
\end{equation}

Closure traces are a generalization of the classical closure amplitude and phase (i.e., the latter can be understood as special cases of the former), but with the important addition of being sensitive to the source polarization structure, in a way independent of virtually any instrumental polarization. Closure traces encode information of the differential polarization throughout the structure of the source. These unique properties of the closure traces serve to testify the latent potential of these observables to complement and improve the analysis of the polarization structure of distant radio sources. However, even though \citet{Broderick&Pesce2020} define the closure traces and explore their symmetries and degeneracies, being particularly relevant the demonstration on the insensitivity of the closure traces to arbitrary coherent (i.e., global throughout the source) rotations in the Poincare sphere, a thorough analysis of the potentially codified information within the closure traces of interferometric observations has yet to be done.

In this paper, we apply the newly-introduced closure traces to extract robust information on the polarization structure of the sources and their possible dependence on the observation frequency. The importance of this study lies within the fact that, with a method for obtaining the information of the source polarimetry (independent of the instrumental polarization) hidden in the closure traces, we could analyse this information and scientifically explore polarization observations where the parallactic-angle coverage of the correlator is insufficient and, as a result, a proper calibration of the data is not possible.

We organize this paper as follows. After providing in Section \ref{Section2} a detailed description of both simulated and observational data used, we present in Section \ref{Section3} a method to extract polarimetric information from observed radio sources by fitting parametric models to closure traces. We test this method on realistic simulations from interferometric observations of the M87 Active Galactic Nucleus (AGN) with the ALMA telescope and real data taken with the ALMA telescope during the EHT observations in the April 2017 campaign. Finally, we show in Section \ref{Section4} the results of our test and discuss the possibility of using closure traces to determine Faraday Rotation (FR) from broadband interferometric observations.

\section{Observations and simulated data}
\label{Section2}
The ALMA radio telescope \citep[see e.g.][]{ALMAWootten2009,ALMAPartnership2015}  is the most powerful and sensitive telescope in the mm/submm wavelength regime, and it is, consequently, a key component of the EHT. This paper concerns, as a test dataset, the ALMA observations conducted on the April 2017 EHT campaign \citep[see][]{EHTIcomplete}. On this campaign, besides recording the signal from the observations in a VLBI backend for its subsequent processing with the rest of the EHT stations, the ALMA correlator computed all the visibilities among the phased antennas. 

We use the ALMA data obtained from the observations of the M87 AGN conducted on April 2017, and properly calibrated by the ALMA observatory. To observe this source, ALMA used an array of 33 antennas, with 12-meter diameters each, with unprojected baseline lengths ranging from $15.1$\,m to $160.7$\,m. The observations were performed in the ALMA Band 6 (ranging from $212$\,GHz to $230$\,GHz). The ALMA heterodyne receivers can sample the radio signal in various parts of the spectrum simultaneously; specifically, the spectrum accessible to ALMA in one observation is separated into two sidebands. Each of these sidebands has two spectral windows (spws) of $2$\,GHz each. The measurements were taken in four different spws ($0$, $1$, $2$ and $3$, corresponding to frequencies centered at $213.1$\,GHz, $215.1$\,GHz, $227.1$\,GHz and $229.1$\,GHz, respectively)  for all epochs of the EHT campaign \citep[see][for details on the setup of these observations]{Goddi2019}.

\begin{figure}[!ht]
    \centering
    \resizebox{\hsize}{!}{\includegraphics{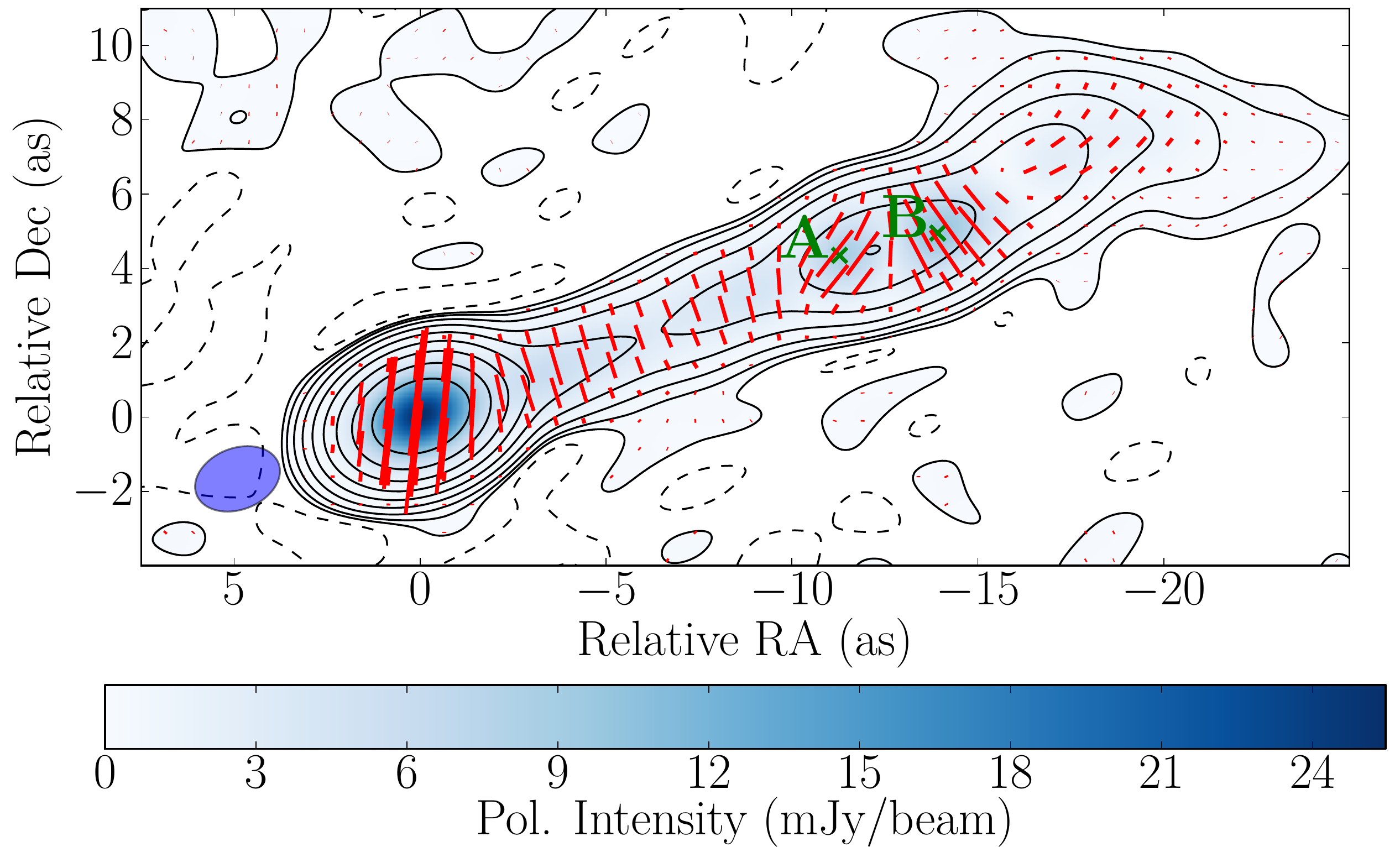}}
    \caption{Image of the polarization structure of the AGN in galaxy M87. The blue scale corresponds to the mapping of the linear polarization intensity. The red lines show the EVPA, with lengths proportional to the polarization intensity. The contours quantify the total intensity (Stokes I). The contours are spaced logarithmically between the image peak (1.21\,Jy/beam) and 0.325\,mJy/beam, which corresponds to the root-mean-squared (rms) of the image residuals. The dashed contour marks the first negative contour in total intensity, at -0.325\,mJy/beam. The full width at half maximum (FWHM) of the convolving Gaussian beam is shown at the bottom-left corner. An alternative version of this figure can be found in \citet{Goddi2021complete}.}
    \label{fig:fig1}
\end{figure}

In this paper, we focus on the M87* track observed on 2017 April 6, between 00:52 and 08:02 UT \citep[see][for details]{Goddi2021complete}. Figure \ref{fig:fig1} shows the polarization structure of the AGN and the jet of M87, obtained from the ALMA data \citep[see][]{Goddi2021complete}.

\begin{figure}[!ht]
    \centering
    \resizebox{\hsize}{!}{\includegraphics{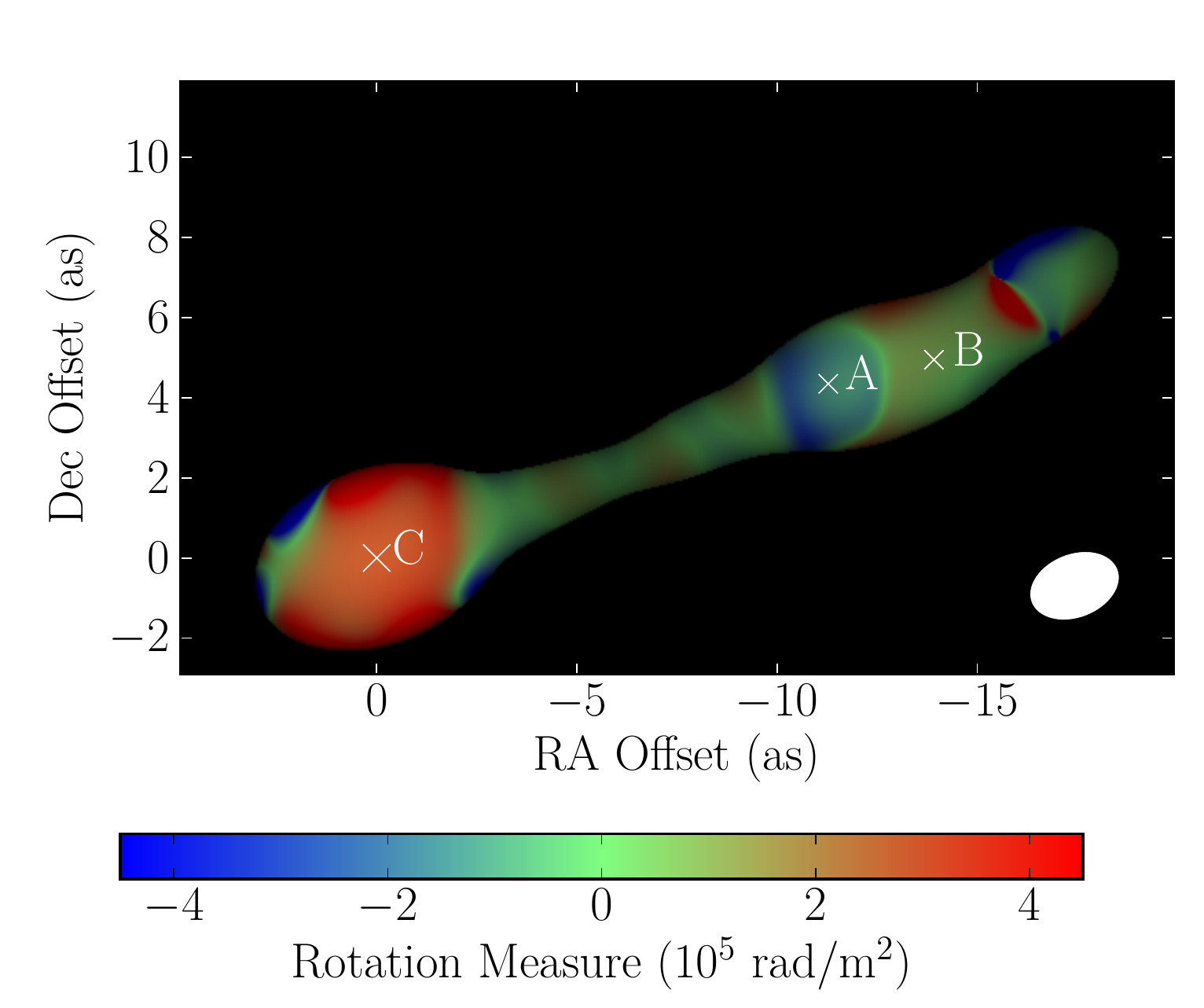}}
    \caption{Rotation Measure image of the M87 galaxy detected from the ALMA observations analysed in this paper. The full width at half maximum (FWHM) of the convolving Gaussian beam is shown at the bottom-right corner.}
    \label{fig:fig2}
\end{figure}

From the brightness distribution of polarized intensity shown in Fig. \ref{fig:fig1}, both the active core of M87 and the two peaks of polarization intensity (knots $A$ and $B$) can be distinguished. Moreover, the Rotation Measure (RM) analysis of the M87 ALMA observations described in \cite{Goddi2021complete} shows a clear detection of FR from the region of the M87 core. On the other hand, the RM synthesis performed across the whole source brightness distribution shows a clear {\em differential} FR between the core and the two jet knots, $A$ and $B$, being the RM at the core notably larger than those at the knots. In Fig. \ref{fig:fig2}, we show the RM map of M87, obtained from the results presented in \cite{TFG}.

\subsection{Simulations and model parameters}
\label{subsec:simulation}

In order to investigate how the polarization structure of a M87-like source (and its dependence with frequency) may affect the closure traces, we generate synthetic data of a source with the more relevant components of M87 (i.e. a core and two peaks of polarization intensity) based on the exact baseline coordinates of the observations described in the previous section.

\begin{table*}[!ht]
\centering
\caption{Simulated model component positions and Stokes values.}
\label{tab:simulM87_data}
\begin{tabular}{c c c c c c c c }
\hline\hline
M87 components         &      $l$     &       $m$                    & spw & I (Jy) & Q (Jy) & U (Jy) & V (Jy) \\
\hline
                                 &                          &                        & 0   & 1.2        & +0.1     & $\sqrt{3}/10$     & 0.     \\
                                 &                          &                        & 1   & 1.2        & +0.3     & $\sqrt{3}/10$     & 0.     \\
                                 &                          &                        & 2   & 1.2        & +0.3     & $-\sqrt{3}/10$     & 0.     \\ 
\multirow{-4}{*}{core}           & \multirow{-4}{*}{0}      & \multirow{-4}{*}{0}    & 3   & 1.2        & +0.1     &  $-\sqrt{3}/10$     & 0.     \\
\hline
                                 &                          &                        & 0   & 0.1        & +0.0018     & $-$0.0081     & 0.     \\
                                 &                          &                        & 1   & 0.1        & +0.0018     & $-$0.0079     & 0.     \\ 
                                 &                          &                        & 2   & 0.1        & +0.0016     & $-$0.0074     & 0.     \\
\multirow{-4}{*}{knot A}         & \multirow{-4}{*}{-11.28} & \multirow{-4}{*}{4.35} & 3   & 0.1        & +0.0015     & $-$0.0072     & 0.     \\
\hline
                                 &                          &                        & 0   & 0.05        & +0.0037     & +0.0090     & 0.     \\
                                 &                          &                        & 1   & 0.05        & +0.0037     & +0.0088     & 0.     \\ 
                                 &                          &                        & 2   & 0.05        & +0.0033     & +0.0076     & 0.     \\ 
\multirow{-4}{*}{knot B}         & \multirow{-4}{*}{-13.92} & \multirow{-4}{*}{4.95} & 3   & 0.05        & +0.0032     & +0.0075     & 0.     \\ \hline
\end{tabular}
\end{table*}

For the simulation of closure traces related to a simplified model of M87 at ALMA scales, we parameterize the polarization intensity as a set of compact (delta) components, associated to the most prominent polarization features of the true image: the core of M87 and the two polarization knots, $A$ and $B$. The Stokes parameters of each of these components were set to different values for each spectral window (i.e., for each frequency), to simulate the effects of FR in the M87 visibilities. The component positions and Stokes values are given in Table \ref{tab:simulM87_data}. The details of the simulation procedure are described below.

The simulated brightness distribution of Stokes $I$ for each component is taken from the values given by the deconvolved model obtained from the real ALMA data at spw $0$. Regarding the rest of Stokes parameters (i.e., $Q$, $U$ and $V$), the model consists of three point sources, located at the positions marked in Fig. \ref{fig:fig2} (which correspond to the jet core and the two knots, $A$ and $B$). The stokes parameters $Q$, $U$ and $V$ of the two jet components are taken from the local peak intensity values of the real full-polarization image (Fig. \ref{fig:fig1}) at each spectral window.

On the other hand, the values $Q$, $U$ and $V$ of the core component are set manually (see Table \ref{tab:simulM87_data}) to simulate a source with greater EVPA rotation compared to M87, thus increasing the RM of our simulated source to $\sim 0.35 \times 10^7$ rad/m$^{2}$, compared to the value of $\sim 1.5 \times 10^5$ rad/m$^{2}$ for M87*, obtained from the ALMA EHT data  \citep[see][]{Goddi2021complete,TFM}. Doing so allows us to test the performance and reliability of our method, since the high simulated RM produces clear imprints in the closure traces.

The simulated data are obtained by computing the discrete Fourier transform of the Stokes parameters ($(\tilde{I},\tilde{Q},\tilde{U},\tilde{V})$), for each $(u,v)$ coordinate and observed frequency $\nu$, that is,
\begin{align}
\label{discreteFTStokes}
    \begin{pmatrix}
      \tilde{I}(u,v;\nu) \\
      \tilde{Q}(u,v;\nu) \\
      \tilde{U}(u,v;\nu) \\
      \tilde{V}(u,v;\nu)
    \end{pmatrix} = \sum_i
    \begin{pmatrix}
      I_i(\nu) \\
      Q_i(\nu) \\
      U_i(\nu) \\
      V_i(\nu)
    \end{pmatrix} exp\left(2\pi j (u_\lambda l_i + v_\lambda m_i) \right),
\end{align}
where $u_\lambda$ and $v_\lambda$ are the $(u,v)$ coordinates in units of wavelength,  $(I_i,Q_i,U_i,V_i)$ are the Stokes parameters and $l_i$, $m_i$ are the direction cosines \citep[see e.g.][]{thompson2017interferometry} measured with respect to the axes $u$ and $v$, that is, the angular offsets with respect the phase center (i.e. the core of M87), in the tangent plane along the right ascension and declination axis, of each of the components of the source.

To finish off our simulation, we introduce, at each integration time, an uncorrelated Gaussian noise contribution to the real and imaginary parts of the visibilities, with zero average and a standard deviation of 0.5\,mJy (similar to the average noise levels per integration time seen in the real data).

\subsection{Prior discussion on the behaviour of the closure traces}
\begin{figure*}[!ht]
\centering
   \includegraphics[width=18cm]{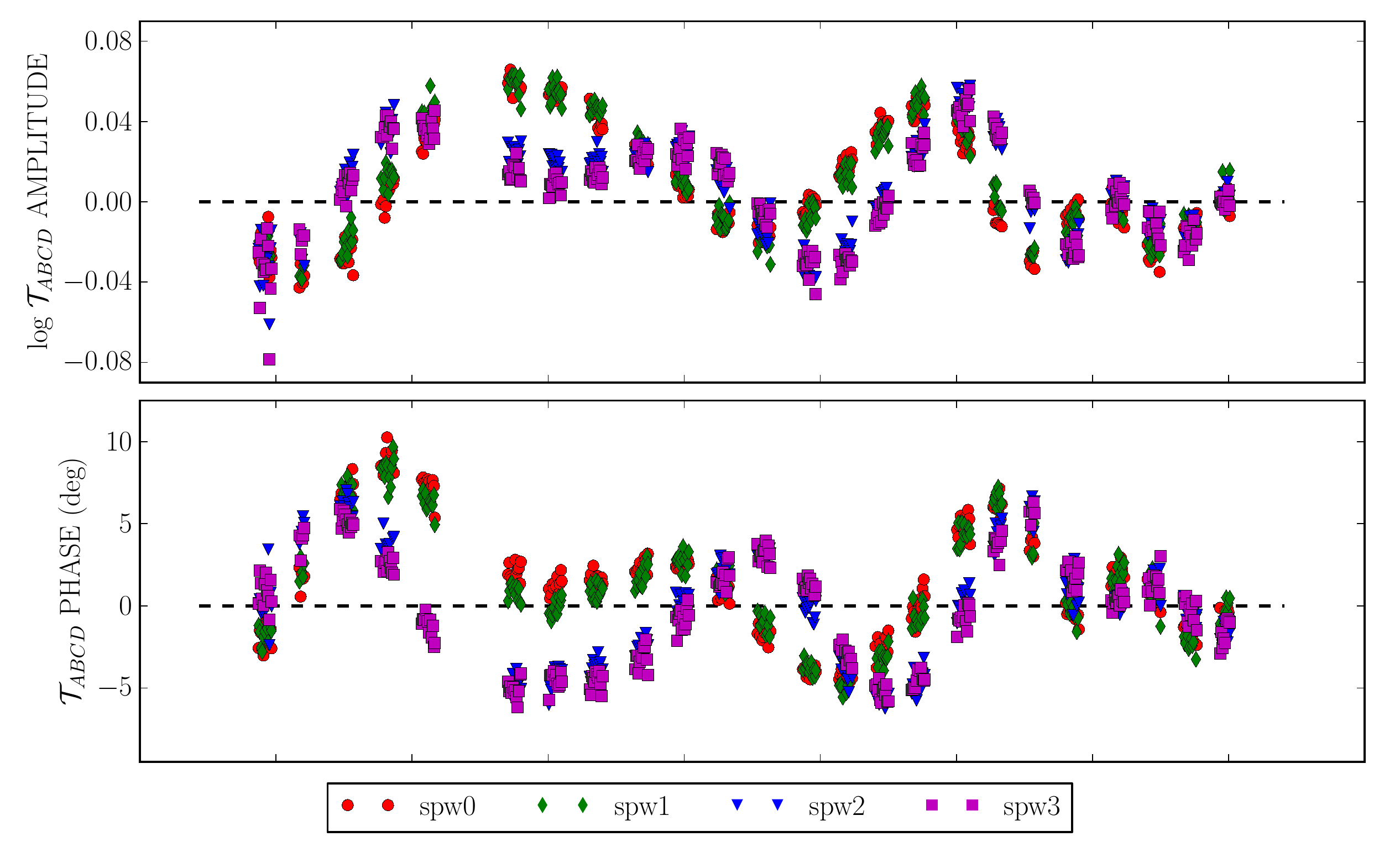}
    \caption{Closure trace phases and (logarithmic) amplitudes for the observational dataset from the ALMA observations of the M87 AGN, for each observed frequency (spws $\{0;1;2;3\}$), for the quadruple of antennas DA50-DA52-DV05-DV12.}
    \label{fig:fig3}
\end{figure*}

Before exploring the $\chi^2$ parametric space with different methods using the closure traces (as proposed in section \ref{Section3}), we analyse the behaviour of the traces for $4$ ALMA stations (e.g. DA50-DA52-DV05-DV12, see Fig. \ref{fig:fig3}) for the ALMA observations of the M87 AGN. As the figure shows, both the closure-trace amplitude and phase behaviour differ for the different spws. This is blatant when comparing the results for the two sidebands of the spectrum explored by ALMA (i.e. spws $\{0,1\}$ versus $\{2,3\}$).

In short, the closure traces behaviour shows a variation of the amplitude and phase with the observed frequency. This variation can be partially attributed to the FR, since it is, by definition, the magneto-optical phenomenon characterized by the variation of the polarization angle with frequency. However, another not-so-obvious cause is the structure factor of the source (i.e., the radial sampling of Fourier space by a baseline, related to the baseline's re-projection in multi-frequency observations). At higher observing frequencies, visibilities are observed at greater distances within the Fourier plane, resulting in a slightly different UV coordinate where the Fourier transform is being measured and, namely, in a slightly different closure trace. Both factors contribute, so this visible appreciation of the variation in the behaviour of the closure traces with frequency might be justified even if the source FR was low. To check the importance of source structure in the spectral dependence of the closure traces, we compare the \textit{model} visibilities, computed from the Stokes $I$ deconvolved model obtained from the real ALMA data, using Eq. \ref{discreteFTStokes} at $213.1$\,GHz and $229.1$\,GHz (i.e. spws $0$ and $3$). We retrieve an average deviation of the \textit{model} visibilities at spws $0$ and $3$ of $(1.2\pm0.7)\%$, much lower than the average deviation computed directly from the visibilities, of $(10\pm20)\%$. Therefore, the frequency dependence seen in the observed closure traces of M87 may be related to FR.

\section{Methodology}
\label{Section3}

Our goal is to study how closure traces can be used to determine the polarimetric structure of sources from interferometric data, that is, we want to quantify the spatial variation of the Stokes parameters across the source brightness distribution. Thus, we need to define a model of the Stokes parameters and compute an error function that can be used to fit the model to the closure-trace data.

Our final objective is to obtain the parameter values of our source polarization model by model-fitting the closure traces only, that is, by searching the set of parameters that minimizes the differences between the closure traces of the simulated or observational data and the parametrized model. We start this analysis by fitting a source polarization model to the closure traces obtained from the simulated observations, to assess the correct convergence of (and/or the presence of degeneracies in) the closure-based fitting. Then, we can proceed to the analysis of the observational data.

First, we develop a Python module to compute the closure traces of both simulated and model data. This Python module gets the visibilities measured on all the baselines connecting four stations $\{A,B,C,D\}$ and computes the corresponding closure traces $\mathcal{T}_{ABCD}$, defined in Eq. \ref{closure_trace}. The error function used in the least-squares minimization is simply given by the squared differences between the data closure traces and those from the parameterized model:
\begin{equation}
\label{chi2_model-fitting}
\chi^2 = \sum_{\{A,B,C,D\}} \abs{\mathcal{T}_{ABCD}^{model}-\mathcal{T}_{ABCD}^{data}}^2.
\end{equation}

We notice, though, that the behaviour of this $\chi^2$ is highly non-linear in the parameter space of the chosen model, with many local minima. Therefore, methods based on the gradient \citep[like the Newton method or the more elaborated Levenberg-Marquardt; see e.g.,][]{LevenbergMarquardt1978} may not converge to the global minimum, depending on the initial parameter values, and the use of methods that explore the complete parametric space, such as Markov chain Monte Carlo \citep[see][]{Goodman&Weare2010,ForemanMackeyMCMC2013,Bussmann2015} or a grid search method, are necessary to analyse more complex sources and obtain better results.

\begin{figure}[!ht]
\centering
\resizebox{\hsize}{!}{\includegraphics{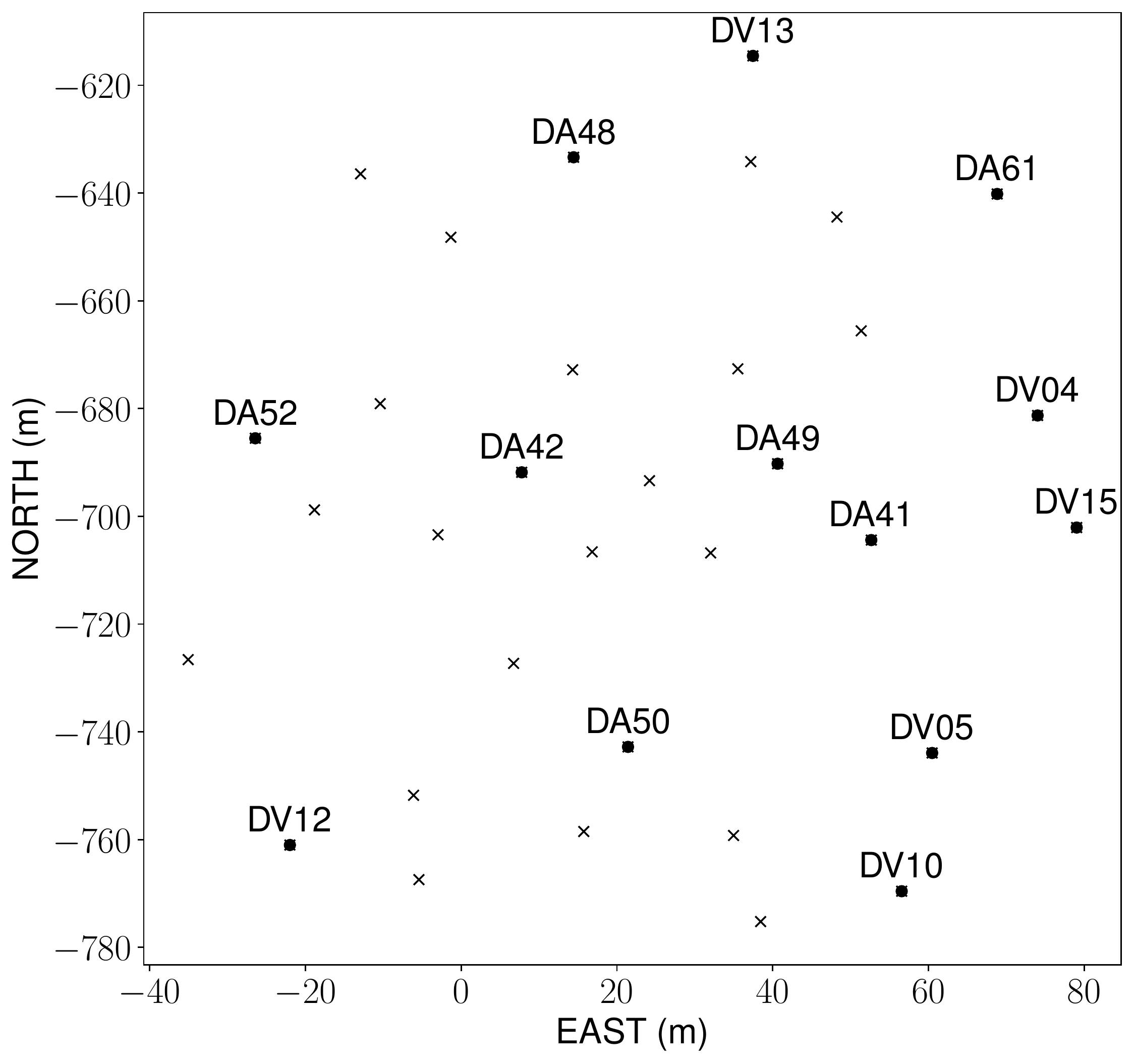}}
\caption{Array of ALMA antennas used in the observations conducted on the April 2017 EHT campaign. The antennas selected for computing the closure traces are highlighted.}
\label{fig:fig4}
\end{figure}

In VLBI observations, typically taken with arrays of around 10 antennas, these methods would be useful to explore the entire parametric space of the $\chi^2$ \citep[see e.g.][]{ThemisAvery2020,PesceDMC2021}. However, for a large number of antennas, which is the case of the ALMA observations analysed here, the computational cost may be prohibitive (for a number of $n$ antennas, there are $n!/\left[4!(n-4)!\right]$ 4-antenna baselines). Therefore, in this work we have decided to use a selection of 13 antennas, distributed in such a way that there is not a significant loss of information in the Fourier space (see Fig. \ref{fig:fig4}).

In the next subsections, we discuss about the use of two different polarization models: one that only accounts for the core differential polarization and a more complete one that also solves for the polarization in the jet. Each model is fitted using different approaches.

\subsection{Modelling the core-only polarization.}
The fitting model consists of a simple parameterization of the source structure, where all the Stokes parameters of the jet components are fixed to the measured values in the ALMA image, collected in Table \ref{tab:simulM87_data}, and the only free parameters are related to the core. Since the closure traces are only sensitive to differential (i.e., contrast) changes within the source (in both total intensity and polarization), we have to parameterize the core's Stokes parameters in a way that encodes differences with respect to the fixed jet components. Otherwise, degeneracies may appear in the fitted parameters. The Stokes parameters of the core are thus formulated using the quantities of the jet components in the following way:
\begin{subequations}
\begin{align}
& Q_{core} = p_1 \ cos\left[2\left(p_0 + \left(\lambda_i^2-\lambda_0^2\right) p_2\right)\right],\\
& U_{core} = p_1 \ sin\left[2\left(p_0 + \left(\lambda_i^2-\lambda_0^2\right) p_2\right)\right],
\end{align}
\label{modelfitting_eqs}
\end{subequations}
\noindent where $\lambda_i$ is the wavelength of the spectral window $i$, $\lambda_0$ is the reference wavelength, and $(p_0,p_1,p_2)$ is the set of parameters needed to characterize the polarization of the core. $p_0$ is the polarization angle (the EVPA), $p_1$ is the fractional polarization in units of the flux density of the core (i.e. $p_1 = m \cdot I_{core}$, where $m$ is the fractional polarization) in Jy, and $p_2$ is the RM of the core in units of rad/m$^{2}$, which accounts for the rotation of the polarization angle with the wavelength (i.e. across the observed bandwidth, divided in 4 spws).

We fit each spectral window independently by setting the RM to 0, to avoid a direct fit of the RM (which would increase the dimension of the parameter space), reducing the computational cost. Thus, the new model (fitted to each spectral window) is given by Eq. \ref{modelfitting_eqs} with the parameter $p_2$ fixed to 0 (i.e. there is no RM) as we analyse each spectral window separately. Since this core-only model only has two fitting parameters, we have opted to perform a grid search method (i.e. the 2-dimensional exploration of $\chi^2$) focusing on two parameters, the EVPA and the fractional polarization of the core. This method allows us to visualize and understand the behaviour of the $\chi^2$ for a simple model in a small parametric space. To retrieve the RM, we need to perform a linear fitting of the polarization angle $\phi = \phi_0 + RM \lambda^2$ as a function of the observation wavelength $\lambda$ (i.e. for each spw).

We explore both parameters $(p_0,p_1)$, that is, different values of the EVPA and the fractional polarization of the core (in units of the core flux density, i.e. $p_1=m \cdot I_{core}$), for each spectral window, obtaining different $\chi^2$ values, computed with Eq. \ref{chi2_model-fitting}, for each $(p_0,p_1)$ values and for each observed frequency. This procedure provides 2D grids where the $\chi^2$ is shown, adjusted to the traces, as a function of the linear polarization intensity and the EVPA.

From the distribution of the residual closure traces at the minimum of the $\chi^2$, it is possible to estimate the correct \textit{temperature} of the $\chi^2$ \citep[e.g.][]{Craiu2014} that reflects the natural scatter of the data. The probability density is then recovered as 
$$\mathcal{P}_{\chi^2}(m,\phi) \propto \exp{\left( -\frac{\chi^2}{2T} \right)},$$
\noindent where $T$ is the standard deviation of the closure trace residuals (evaluated for each dataset and spectral window) at the $\chi^2$ minimum, and assuming the same weights for all visibilities.

\subsection{Model-fitting two M87 components: the MCMC approach}
\label{subsec:MCMCapproach}

To retrieve the RM directly from the closure traces, we need to include it as a new parameter in the fitting model, by using Eq. \ref{modelfitting_eqs}, without fixing $p_2$ to 0. Furthermore, as discussed in Sect. \ref{Section2}, the polarized structure of the M87 jet seen by ALMA presents three relevant components, and therefore our simulation is modelled with three compact components.

Thus, a more accurate modelling would be to allow all polarization components to have their own RM. However, the closure traces are only sensitive to differential polarization (or polarization contrast) across the source structure, which means that we must leave one of the components fixed (and force its Stokes $V$ to zero, in order to break the latitudinal degeneracy in the Poincar\'e Sphere) and allow the EVPAs and intensity of the other two components to vary. 

A greater number of parameters for our fitting model increases the computational cost of our grid-search approach. Hence, once studied the grid-search algorithm, we can go one step further and apply a \textit{Markov chains Monte Carlo} (MCMC) approach to an extended set of parameters, by including in the fitting model one of the two most polarized jet components. We choose to fix the knot $B$ because it is the jet component with lower RM (see Fig. \ref{fig:fig2}). Therefore, our new fitting model consist of the following parameterization of the source Stokes parameters
\begin{subequations}
\begin{align}
    Q_{core} &= p_1 \ cos\left[2\left(p_0 + \left(\lambda_i^2-\lambda_0^2\right) p_2 \right)\right],\\
    U_{core} &= p_1 \ sin\left[2\left(p_0 + \left(\lambda_i^2-\lambda_0^2\right) p_2 \right)\right],\\
    Q_{knotA} &= p_4 \ cos\left[2\left(p_3 + \left(\lambda_i^2-\lambda_0^2\right) p_5 \right)\right],\\
    U_{knotA} &= p_4 \ sin\left[2\left(p_3 + \left(\lambda_i^2-\lambda_0^2\right) p_5 \right)\right],
\end{align}
\label{eq:MCMC2comp}
\end{subequations}
\noindent where $(p_0,p_1,p_2)$ and $(p_3,p_4,p_5)$ are the sets of parameters needed to characterize the polarization of the core and the knot $A$. $p_0$ and $p_3$ are the polarization angle (the EVPA), and $p_1$ and $p_4$ are the fractional polarization in units of the flux density, of the core and the knot $A$ (i.e. $p_1 = m \cdot I$, where $m$ is the fractional polarization). $p_2$ and $p_5$ are the RM of the core and the knot $A$ in units of rad/m${^{2}}$.

The MCMC method allows us to test if the closure traces can retrieve the polarization of different components of an observed source, simultaneously, since the grid-search method becomes computationally unfeasible for the extended set of parameters. We use the Python tool \texttt{emcee} \citep[see][]{ForemanMackeyMCMC2013} for a affine-invariant MCMC approach to model-fitting, estimating the differential FR among the source components. Furthermore, as this information is obtained by using closure traces, the results will be independent of the instrumental polarization.

In our model-fitting, we set $\lambda_0 = 0$, and therefore the $p_0$ value to which the MCMC method converges corresponds to the intrinsic polarization angle of the core, $\phi_{0}$, i.e, the EVPA at infinite frequency. If we want to obtain, from this result, the EVPA of the core at another reference frequency (e.g. at the average wavelength $\bar{\lambda}$ of the 4 spws observed by ALMA, $\bar{\phi}$), we need to add the neglected FR effect,
$$ \bar{\phi} = \phi_0 + RM \ \bar{\lambda}^2.$$
We apply this correction to compare the polarization angle to which the MCMC method converges and the EVPA obtained from the image analysis.

An important property the Stokes components $(Q,U)$ parameterization in Eq. \ref{eq:MCMC2comp} is that the polarization angle is degenerated by a $n\pi$ phase, which could hinder the convergence of the MCMC method and give results affected by said degeneracy. We can break this ambiguity using circular statistics \citep[see e.g.][]{Fisher1987Spherical,Jupp2009DirectionalStat}, a simple, but effective, change of parameters of the fitting model, which uses the angle sum identity on Eq. \ref{eq:MCMC2comp},
\begin{align*}
    Q_{core} &= p_1 \cos\left(2p_0\right) \cos\left(2RM'\right) - p_1 \sin\left(2p_0\right) \sin\left(2RM'\right),\\
    U_{core} &= p_1 \sin\left(2p_0\right) \cos\left(2RM'\right) + p_1 \cos\left(2p_0\right) \sin\left(2RM'\right),
\end{align*}
to redefine the parameterization model as
\begin{subequations}
\begin{align}
    Q_{core} &= q_0 \cos\left(2RM'\right) - q_1 \sin\left(2RM'\right),\\
    U_{core} &= q_1 \cos\left(2RM'\right) + q_0 \sin\left(2RM'\right)\\
    Q_{knotA} &= \left[q_2 \cos\left(2RM_A'\right) - q_2 \sin\left(2RM_A'\right)\right] \varphi_A,\\
    U_{knotA} &= \left[q_3 \cos\left(2RM_A'\right) + q_3 \sin\left(2RM_A'\right)\right] \varphi_A,
\end{align}
\label{eq:MCMC2comp_circstat}
\end{subequations}
\noindent where $RM' = \left(\lambda_i^2-\lambda_0^2\right) p_4$ and $RM_A' = \left(\lambda_i^2-\lambda_0^2\right) p_5$, and
$$\varphi_A = exp\left[\frac{2\pi}{\lambda_i}j \left(u \cdot \alpha + v \cdot \delta \right)\right]$$
\noindent is the knot $A$ auxiliary phase; the new parameters (to determine with the convergence of the MCMC method) are defined, respect the old parameterization, as 
$$q_0 = p_1 \cos\left(2p_0\right), \ \ \ \ \ \ q_1 = p_1 \sin\left(2p_0\right),$$
$$q_2 = p_4 \cos\left(2p_3\right) \ \text{and} \ \ q_3 = p_4 \sin\left(2p_3\right).$$
and are unaffected by the EVPA ambiguity, as the ambiguity is absorbed within the parameters of these trigonometric functions (for instance, $p_0 + n\pi$, where $n$ is an integer, corresponds to the same $q_0$, for a given $p_1$). Consequently, we can retrieve the original parameters without the ambiguities after finding the $q_i$ values to which the MCMC method converges,
\begin{subequations}
\begin{align}
    & p_0 = \frac{1}{2} \arctan\left(\frac{q_1}{q_0}\right); \ \ \ \ p_1 = \sqrt{q_0^2+q_1^2}\ ; \\
    & p_3 = \frac{1}{2} \arctan\left(\frac{q_3}{q_2}\right); \ \ \ \ p_4 = \sqrt{q_2^2+q_3^2}\ .
\end{align}
\label{eq:reparametrization}
\end{subequations}
\noindent
Finally, $p_2$ and $p_5$ are preserved in our parameterization as the RM of the core and the knot $A$.

\section{Results and discussion}
\label{Section4}

\subsection{Simulated data}

\subsubsection{Grid Search}

The exploration of the parametric space of the $\chi^2$ using the grid search method (described in Section \ref{Section3}) provides the value of the polarization angle $\phi$ and the linear polarization intensity $m$ that best fit the closure-trace data (hence, independent of calibration) at each spw. We have, thus, one independent set of polarization parameters per spw, which we can compare to the true Stokes parameters used in the simulation.

\begin{table}[h!]
\centering
\caption{Polarization angle, $\phi$, and fractional polarization, $m$, values for which we obtain the minimum of the $\chi^2$ after exploring its parametric space with the grid search method, being $\phi$ and $m$ the parameters $p_0$ and $p_1$ in Eq. \ref{modelfitting_eqs}, respectively, for each of the observed spws. The values used in the simulation are given for comparison.}
\label{tab:grid-search_results}
\begin{tabular}{c cc cc}
\hline \hline
 &  \multicolumn{2}{c}{$\phi \ (deg.)$} &  \multicolumn{2}{c}{$m\cdot I$ (Jy)} \\
\multirow{-2}{*}{spw} &  \multicolumn{1}{c}{grid-search} &  simulated &  \multicolumn{1}{c}{grid-search} &  simulated \\
\hline
0 & \multicolumn{1}{c}{30.11 $\pm$ 0.05}   & 30  & \multicolumn{1}{c}{0.200 $\pm$ 0.001} & 0.200 \\
1 & \multicolumn{1}{c}{14.37 $\pm$ 0.05}   & 15  & \multicolumn{1}{c}{0.347 $\pm$ 0.001} & 0.346 \\
2 & \multicolumn{1}{c}{-13.66 $\pm$ 0.11}  & -15 & \multicolumn{1}{c}{0.348 $\pm$ 0.002} & 0.346 \\
3 & \multicolumn{1}{c}{-25.2 $\pm$ 0.3} & -30 & \multicolumn{1}{c}{0.203 $\pm$ 0.005} & 0.200 \\
\hline
\end{tabular}
\end{table}

\begin{figure*}
    \centering
    \includegraphics[width=18cm]{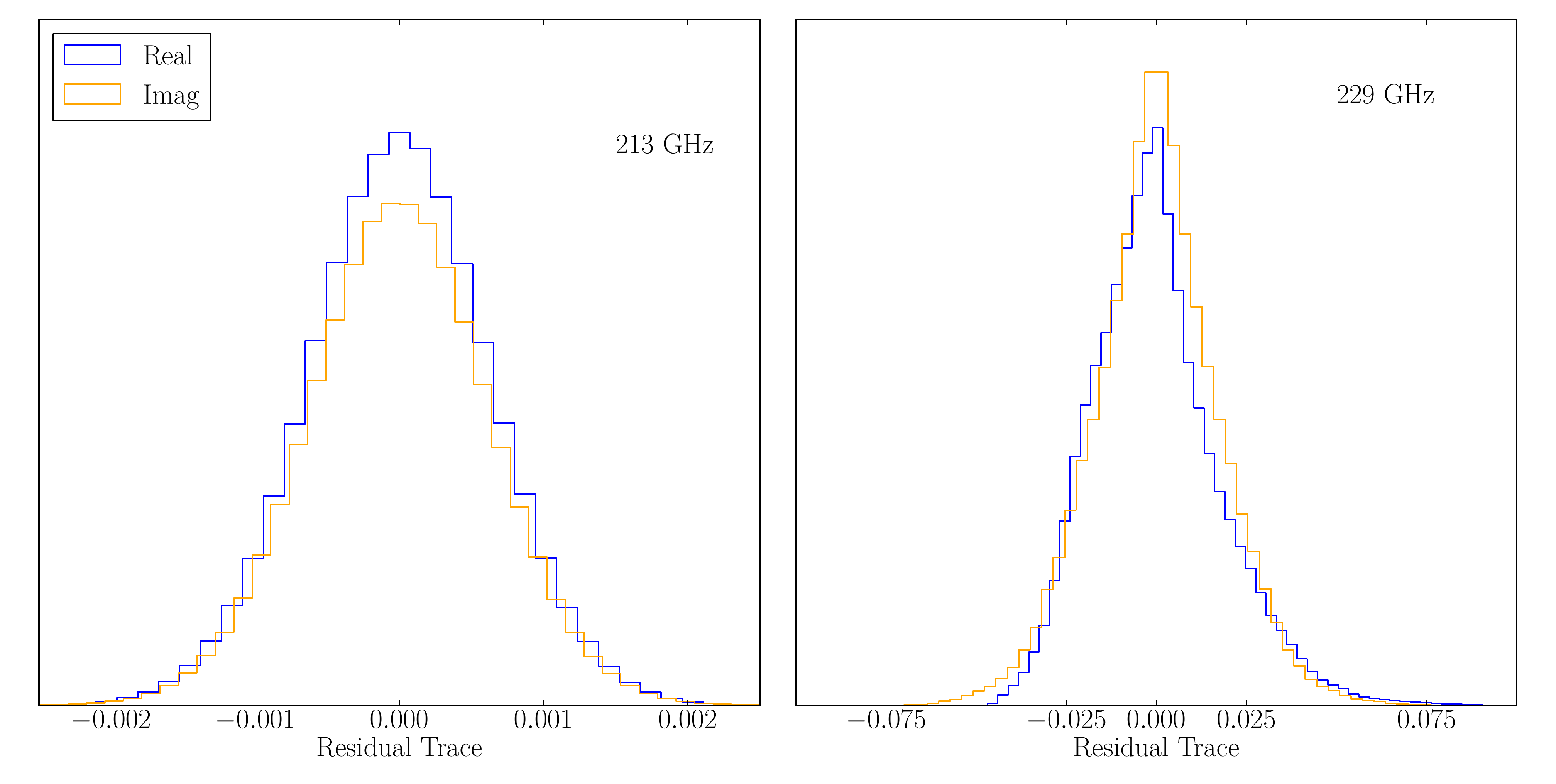}
     \caption{Histograms of the distribution of residual traces at the $\chi^2$ minimum obtained from the exploration of the $\chi^2$ parametric space carried out with the grid search method, obtained with the simulated data for spws $0$ (left panel) and $3$ (right panel).}
     \label{fig:fig5}
\end{figure*}

In Table \ref{tab:grid-search_results}, we collect the result of our (grid-search) exploration of the parametric space of the $\chi^2$ for the $4$ spws  of the simulated data. We obtain a clear dependence of the polarization angle of the core with the spectral window. We also notice that the fitting model, based only in closure traces, retrieves with minor discrepancies the expected EVPA and fractional polarization for spws $0$ and $1$. On the other hand, even though the results from spws $2$ and $3$, which correspond to the other sideband of the separated ALMA spectrum, present greater discrepancies compared to the first sideband, especially for the polarization angle, they are still close to the expected values of our simulation. Fig. \ref{fig:fig5} shows the distribution of residual traces at the $\chi^2$ minimum, that is, at the two fitting parameters of the core (i.e., the EVPA, $\phi$, and the fractional polarization, $m$) which minimizes Eq. \ref{chi2_model-fitting} for our fitting model, for spws $0$ and $3$.

\begin{figure}[!h]
    \centering
    \resizebox{\hsize}{!}{\includegraphics{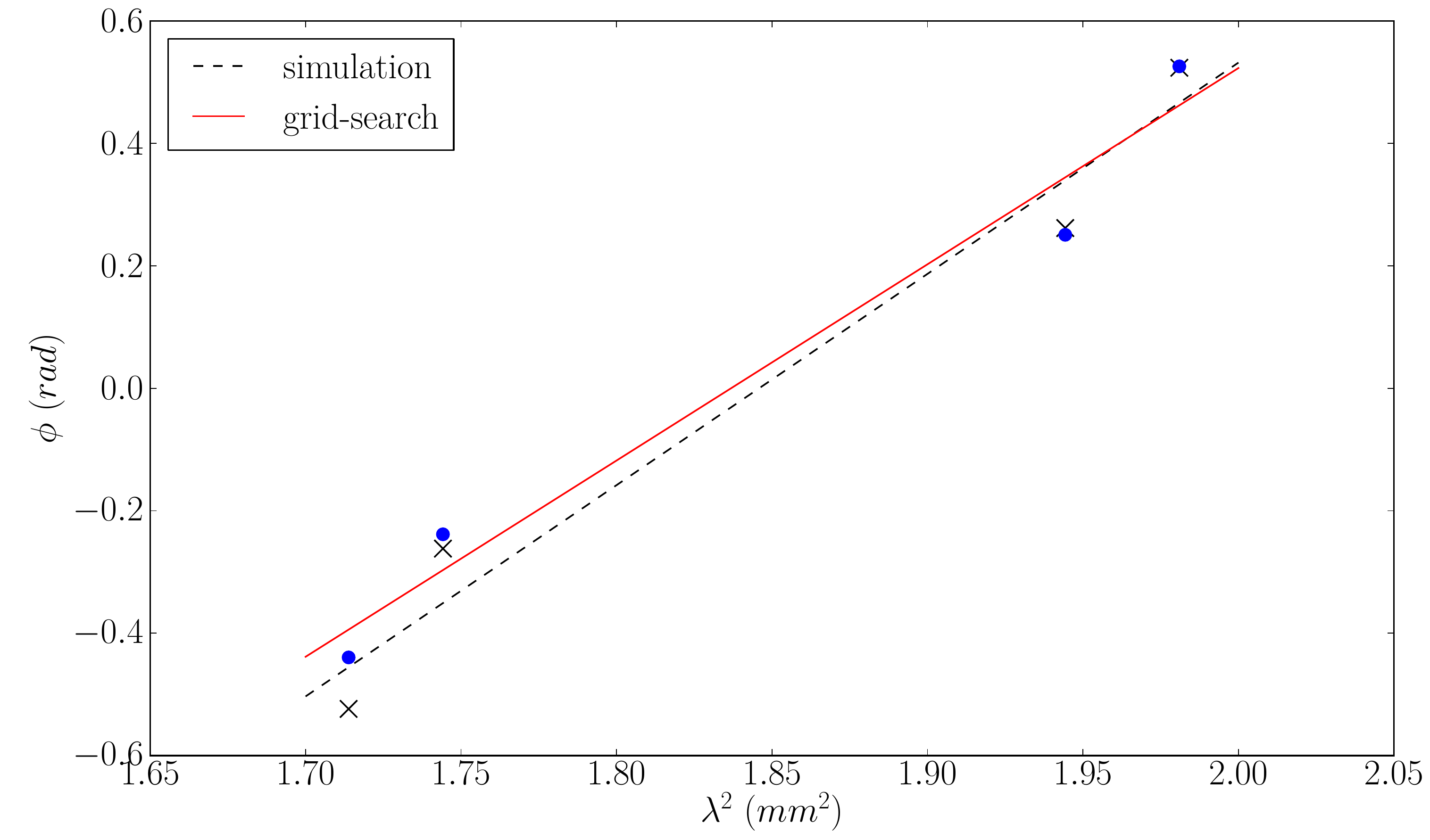}}
    \caption{Linear fitting of the polarization angle of the core, $\phi = \phi_0 + RM \lambda^2$, as a function of the observation wavelength squared, $\lambda^2$. The $\phi$ values are determined from the grid-search exploration of the $\chi^2$ parametric space, by fitting the closure-trace model to the simulated data, with the polarization of knots $A$ and $B$ fixed to their values at spw0 (see Sect. \ref{subsec:simulation}). We obtain a fitted value of $RM = (0.32 \pm 0.04) \times 10^7$ rad/m$^{2}$, compatible with the expected value of $(0.35 \pm 0.04) \times 10^7$ rad/m$^{2}$ obtained from fitting the simulated polarization angles.
}
    \label{fig:fig6}
\end{figure}

From these results, we retrieve the \textit{differential} FR along the core-jet structure of the simulation by fitting the polarization angle as a function of the observation wavelength $\phi(\lambda)$. By doing so, we obtain a value of the RM of the core component, $RM = (0.32 \pm 0.04) \times 10^7$ rad/m$^{2}$, as shown in Fig. \ref{fig:fig6}, which is compatible with the expected value of $(0.35 \pm 0.04) \times 10^7$ rad/m$^{2}$. The expected value is computed by fitting the polarization angle of the simulation, determined from the Stokes values of the core given in Table \ref{tab:simulM87_data}, as a function of the observation wavelength $\phi(\lambda)$ (see Fig. \ref{fig:fig6}).

\begin{figure*}
    \centering
    \includegraphics[width=18cm]{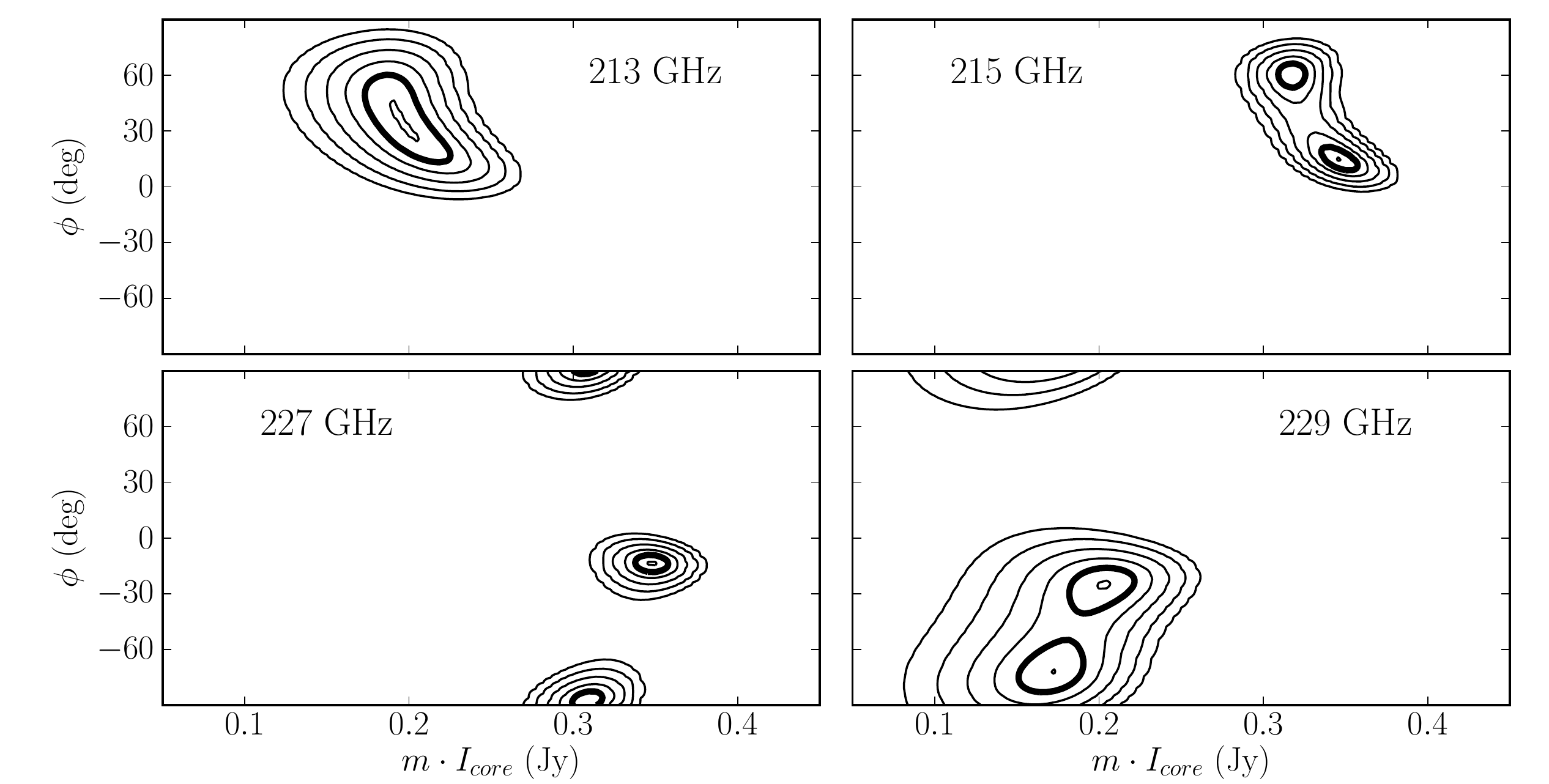}
\caption{Results of the exploration of the $\chi^2$ parametric space carried out with the grid search method, for each of the observed spectral windows. The probability density distribution of the two core fitting parameters (i.e., the EVPA, $\phi$, and the fractional polarization, $m$), corresponding to the $\chi^2$ scaled by the proper temperature, is shown with six contours located (in peak units) at 0.95, 0.317 (i.e., 1$\sigma$), 0.045 (2$\sigma$), 0.0027 (3$\sigma$), 4$\sigma$ and 5$\sigma$. Notice the 180 degrees ambiguity of the $\chi^2$ distribution in the direction of the EVPA.}
\label{fig:fig7}
\end{figure*}

Furthermore, if we explore the complete $\chi^2$ parametric space, as shown in Fig. \ref{fig:fig7}, we notice the closure traces degeneracies when exploring each spw independently, as there are several relative minima of the $\chi^2$ for different values of the polarization angle, $\phi$, and the fractional polarization, $m$. Additionally, we notice that the $\chi^2$ presents a periodic behaviour in the $[-\pi/2,\pi/2]$ interval for the polarization angle, as expected.

\subsubsection{Markov chains}

Up till now, we have analysed the results for a model where only the polarization of the core is being fitted for each spw independently, retrieving the polarization angle and the fractional polarization and then determining the RM with a linear fitting (as shown in Fig. \ref{fig:fig6}). We are using a fitting model with a limited number of parameters, as the computational cost of the grid-search approach increases for a high-dimensional parametric space.

However, by using the MCMC approach described in Sect. \ref{subsec:MCMCapproach}, we can not only include a parameter to retrieve the RM of the core directly combining the visibilities observed for all the spws, but also allow all polarization components to vary in the MCMC exploration (leaving only one of the components fixed, as the closure traces are only sensitive to differential polarization across the source structure). Therefore, our new fitting model consist of the parameterization of the source core and knot $B$ Stokes $Q$ and $U$, using the polarization angle, the fractional polarization and the RM of each component as fitting parameters.

\begin{figure}[!h]
\centering
    \resizebox{\hsize}{!}{\includegraphics{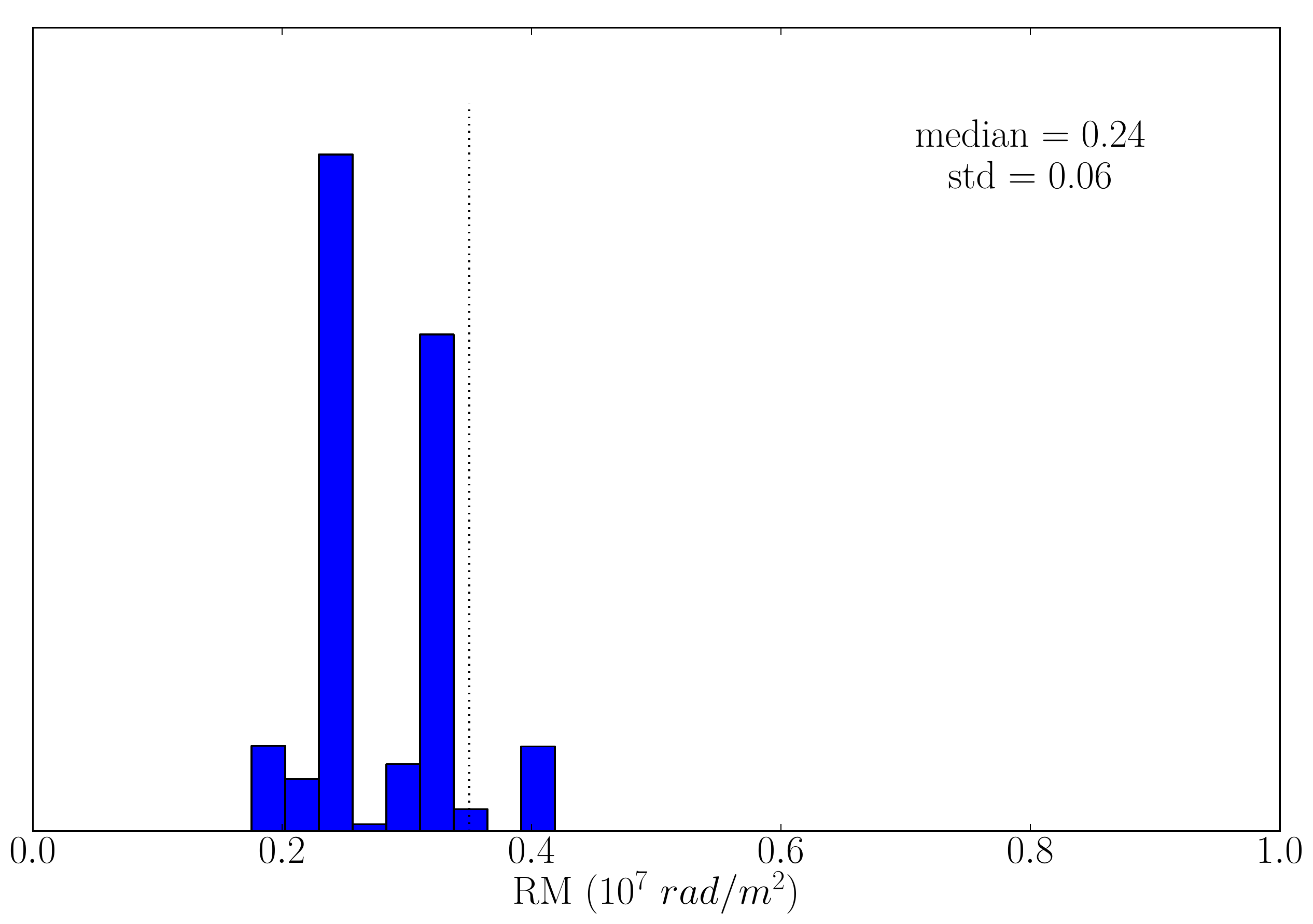}}
     \caption{Results of the MCMC posterior sampling of the model consisting of three polarized components, fitting two of them (the core and knot $B$ of the simulated M87-like source), for the simulated data. The histogram shows the RM of the core, and the dotted line indicates the (higher) expected value, $(0.35 \pm 0.04) \times 10^7$ rad/m$^{2}$, increased in our simulation, from the image analysis.}
     \label{fig:fig8}
\end{figure}

In addition to this, for a correct MCMC exploration of the source polarization, and to avoid the effect of EVPA ambiguities, we make use of the circular statistics parameterization (the fitting model described by Eq. \ref{eq:MCMC2comp_circstat}). In Fig. \ref{fig:fig8}, we present the results of the MCMC model-fitting, once retrieved the polarization angle and the fractional polarization from the circular statistics parameters with Eq. \ref{eq:reparametrization}.

We obtain a RM posterior distribution spread around $RM \sim (0.2-0.4) \times 10^7$ rad/m$^{-2}$, which is close to, albeit a bit lower than, the expected value of $(0.35 \pm 0.04) \times 10^7$ rad/m$^{2}$ (see Fig. \ref{fig:fig6}). This result proves that, even though the $\chi^2$ parametric space presents several minima when model-fitting for each spw individually, as reported in Fig. \ref{fig:fig7}, due to the degeneracies of the closure traces with the EVPA and the fractional polarization, combining all the spw seems to help breaking the degeneracies, as we retrieve an RM posterior with values close to the true one, within a factor 2.

Finally, the discrepancies reported for both methods could be explained by the combined effect of introducing gaussian noise on the simulated visibilities, the use of circular statistics and/or allowing one knot component to vary, and the loss of information from using a subset of 13 antennas in our closure-based modelling (due to the computational limitations), which could also explain the wide width of the RM posterior distribution obtained using the MCMC approach.

\subsection{Real ALMA Observations}
\label{subsec:resultsALMA}

\begin{figure}[h!]
\centering
    \resizebox{\hsize}{!}{\includegraphics{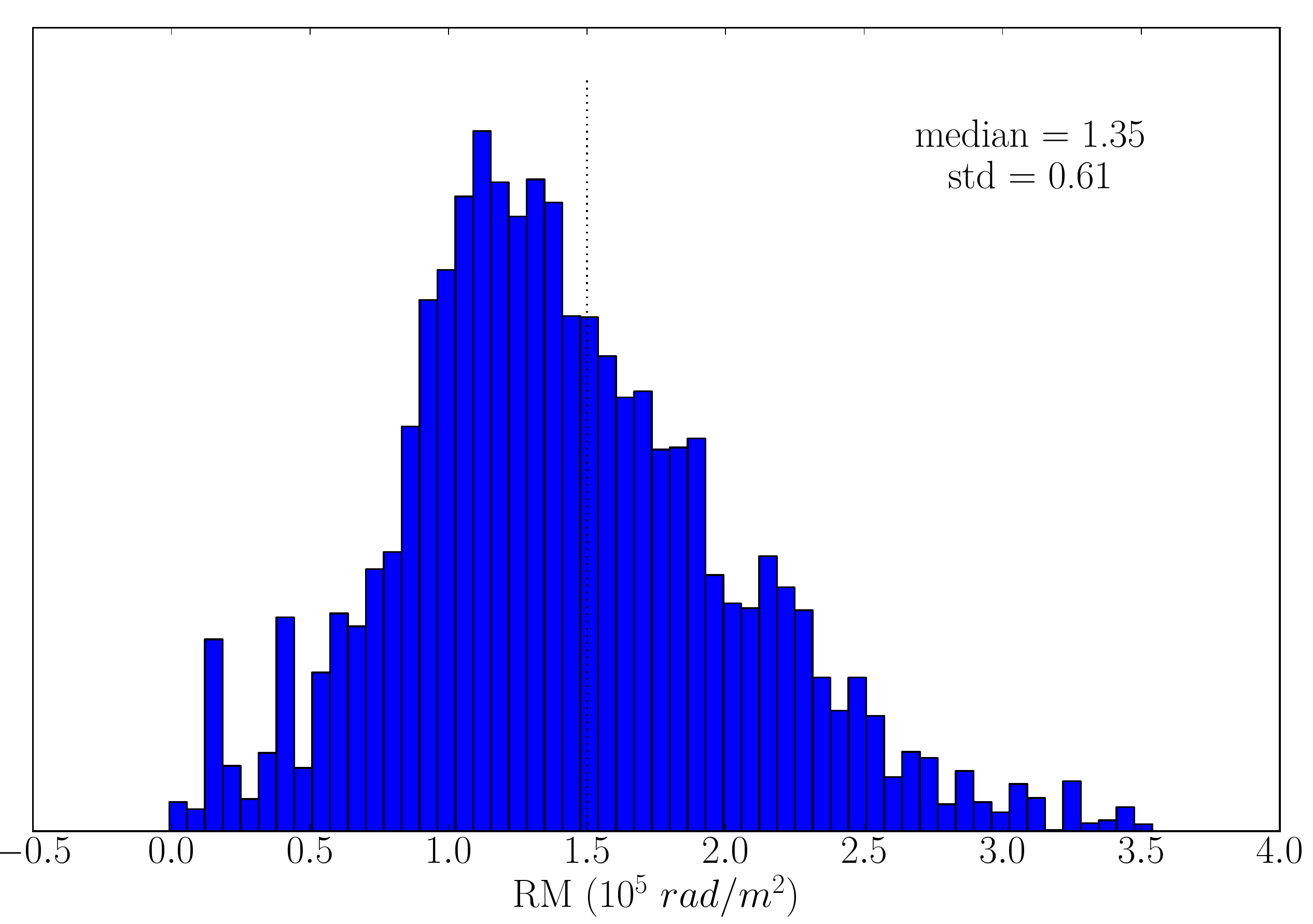}}
     \caption{Results of the MCMC posterior sampling of the model consisting of three polarized components, fitting two of them (the core and knot $A$ of the M87 AGN+jet structure), for the real ALMA observations. The histogram shows the RM of the core, and the dotted line indicates the expected value, of $\sim 1.5 \times 10^5$ rad/m$^{2}$, from the image analysis \citep[see][]{Goddi2021complete,TFM}.}
     \label{fig:fig9}
\end{figure}

We have proven that the degeneracy of the closure traces with the EVPA and the fractional polarization observed for each spw is broken when using the MCMC method to quantify the FR (i.e. to retrieve directly the RM as a parameter of the fitting model). This motivates the last part of our analysis on the use of closure traces on the real ALMA observations of M87 to retrieve the differential FR. The RM posterior distribution obtained from the MCMC model-fitting approach is shown in Fig. \ref{fig:fig9}.

We now compare our results with the results obtained from the image analysis \citep[see][]{Goddi2021complete,TFM}. We get a RM posterior distribution peaked around $RM \sim (0.8-1.6) \times 10^5$ rad/m$^{-2}$, being consistent with the values obtained from the image analysis. Therefore, from these results, we can infer that there is indeed a \textit{differential} FR along the core-jet structure of M87 AGN. However, the width of the RM posterior distribution is very wide (i.e., the precision of the RM is poor), likely due to the small EVPA rotation (of the order of 2 deg.) of the core across the ALMA band. In addition to the small EVPA rotation at the core, we are using only a subset of 13 antennas in our closure-based modelling (due to the computational limitations), which has a negative impact on the achievable signal-to-noise ratio (SNR), as compared to that of the image analysis, where all the phased ALMA antennas have been used. However, for VLBI arrays (consisting typically of 10-20 antennas), it is possible to use the whole interferometer in the closure-based fitting.

If the fractional bandwidth covered by the observations was much larger than that of these ALMA observations \citep[for example, the case of the \textit{VLBI Global Observing System} (VGOS), with practically continuous coverage between 2 and 14 GHz, see][]{Petrachenko2012, Alef2019}, the precision of the RM could be higher. This would make the use of closure traces for imaging VGOS data a promising task.

The results presented in this work, obtained exclusively and directly using closure traces, are independent of the instrumental effects which contaminate the signal, since the closure traces are insensitive to these effects. We have shown that the use of closure traces on simulated and real observations, with the proper selection of fitting parameters, allows to retrieve information of the source polarization. These results pave the way towards new image reconstruction algorithms that could rely on them, possibly improving their performance, as these quantities are insensitive to all sorts of antenna-based and direction-independent calibration effects.

\section{Conclusions}
\label{Conclusions}

Polarization calibration of interferometric observations is a complex procedure prone to different kinds of errors. For the calibration to be successful, strong calibrators, either strongly polarized (case of observations with linear feeds) or ideally weakly polarized (case of observations with circular feeds) need to be done, covering a wide range of parallactic angles for the participating antennas. 

For observations where the parallactic-angle coverage of the correlator is insufficient, a proper calibration of the data is not possible. Consequently, rather strong quality assurance (QA) conditions are imposed to polarization observations, resulting in several observing epochs to be discarded.

In this sense, the discovery of calibration-independent quantities that are sensitive to the source intrinsic polarization \citep[the closure traces, ][]{Broderick&Pesce2020} opens very promising possibilities to develop algorithms able to retrieve robust polarimetric information that will be free from instrumental effects.

In this publication, we use a simple least-squares minimization approach, based on a non-degenerate set of fitting parameters in trace space (i.e., focusing on differential, or contrast, polarization changes across the source structure), to retrieve source-polarization information from a set of ALMA observations. 

We successfully test our method with simulated ALMA data and, once applied to real observations, we are able to recover the RM of the core of M87 (as observed in ALMA Band 6 during the EHT 2017 campaign) using only closure-trace observables.

Our method can be applied to any kind of interferometer, observing in any kind of polarization basis. Possible uses could include, for instance, studies of FR in AGN jets from VLBI observations, from which longitudinal and transversal RM gradients provide important information about the structure and dynamics of the magnetic fields associated to the jet propagation \citep[e.g., see][]{O'Sullivan&Gabuzda2009,O'Sullivan_et_al_2012}.

\begin{acknowledgements}

This work has been supported by the grant PRE2020-092200 funded by MCIN/AEI/ 10.13039/501100011033 and by ESF invest in your future.

This work has been partially supported by the Generalitat Valenciana GenT Project CIDEGENT/2018/021 and by the MICINN Research Project PID2019-108995GB-C22.

We acknowledge support from the Astrophysics and High Energy Physics programme by MCIN, with funding from European Union NextGenerationEU (PRTR-C17I1) and the Generalitat Valenciana through grant ASFAE/2022/018.

This paper makes use of the following ALMA data: ADS/JAO.ALMA\#2016.1.01154.V ALMA is a partnership of ESO (representing its member states), NSF (USA) and NINS (Japan), together with NRC (Canada), MOST and ASIAA (Taiwan), and KASI (Republic of Korea), in cooperation with the Republic of Chile. The Joint ALMA Observatory is operated by ESO, AUI/NRAO and NAOJ.
\end{acknowledgements}

\bibliographystyle{aa}
\bibliography{Albentosa-Ruiz_paper.bib}

\begin{appendix}

\section{Analytical Study of the Closure Traces}
\label{Ap:Section_1.5}

The primary data products obtained by an interferometer are the observed visibilities, as the Radio Interferometer Measurement Equation (RIME) \citep[see][]{Hamaker1996I,Hamaker2000} sets a connection, by means of the Fourier transform, between the observed visibilities and the Stokes maps of astronomical sources. In modern astronomical radio interferometry, one of the greatest difficulties to be solved in order to analyse the data obtained in the observations is to properly characterize the instrumental gain and polarization terms (also known as \textit{D-terms}) that contaminate the visibility matrices, modelled by RIME as Jones matrices.

Traditionally, the closure amplitude and closure phase \citep[see e.g.][respectively]{Jennison1958,Twiss1960} have been used to help with the calibration thanks to their invariance to the antenna dependent atmospheric effects and electronic gains. Even thought they are completely invariant to antenna-based phase errors and atmospherical gains, their usefulness is limited as they are not invariant to non-degenerate calibration errors (e.g. instrumental polarization effects modelled as the D-terms). In this context, \citet{Broderick&Pesce2020} reported a new closure quantity, the closure trace, an interferometric observable characterized for being invariant to the effects of any set of Jones matrices contaminating the visibilities\footnote{For detailed description on the \textit{Radio Interferometer Measurement Equation} (RIME) matrix formalism, which describes the response of a radio interferometer to a signal, \citep[see][]{Hamaker1996I,Hamaker2000}}.

The closure traces are complex closure quantities constructed from the visibility matrices $\pmb{V}$ measured for baselines connecting four antennas $\{A,B,C,D\}$, as follows:
\begin{equation}
\mathcal{T}_{ABCD}=\frac{1}{2} tr(\pmb{V}_{AB}\pmb{V}^{-1}_{CB}\pmb{V}_{CD}\pmb{V}^{-1}_{AD}).
\end{equation}
With this definition, a total of 24 complex traces can be built for the different combinations of possible antennas, with only 6 non-redundant traces \citep[see][]{Broderick&Pesce2020}: $\mathcal{T}_{ABCD}, \mathcal{T}_{ABDC}, \mathcal{T}_{ACBD}, \mathcal{T}_{ACDB}, \mathcal{T}_{ADBC}$ y $\mathcal{T}_{ADCB}$.

It is on interest to understand, at least qualitatively, how differential variations of the Stokes parameters $Q$ and $U$ throughout the source influences the behaviour of the closure traces. To simplify our analysis, we study analytically the closure traces of the double source \citep[see][]{TFM}. 

\subsection{The case of the double source}
Limiting the problem to a 1-spatial dimension, the polarized structure of the double source, given by the Stokes parameters distribution $(I,Q,U,V)$, can be expressed as a sum of Dirac deltas placing the origin of coordinates at the intermediate point of the separation $d$ of the two double source components. The Fourier transform of the Stokes maps $(I,Q,U,V)$ is, evaluated for the spatial frequency $u$, in units of the observed wavelength $\lambda$,
\begin{align}
\label{FTStokes}
    \begin{pmatrix}
      \tilde{I} \\
      \tilde{Q} \\
      \tilde{U} \\
      \tilde{V}
    \end{pmatrix} = 
    \begin{pmatrix}
      I_1 \\
      Q_1 \\
      U_1 \\
      V_1
    \end{pmatrix}   exp\left(i\frac{\pi u}{\lambda}d\right) + 
    \begin{pmatrix}
      I_2 \\
      Q_2 \\
      U_2 \\
      V_2
    \end{pmatrix}   exp\left(-i\frac{\pi u}{\lambda}d\right),
\end{align}
where $(I_i,Q_i,U_i,V_i)$ are the Stokes parameters of the double source components $1$ and $2$.

We study the behaviour of the closure trace calculated from the visibilities obtained in an interferometric observation with 4 stations, $\{A,B,C,D\}$. For circular feeds, the visibilities are related to the Fourier transform of the Stokes parameters, for a baseline $A-B$, via \citep[see e.g.][and references therein]{Smirnov2011},
\begin{equation}
\bar{\pmb{V}}_{AB} = 
\begin{pmatrix}
\tilde{I}(u_{AB}) + \tilde{V}(u_{AB})  & \tilde{Q}(u_{AB})  + i\tilde{U}(u_{AB})  \\
\tilde{Q}(u_{AB})  - i\tilde{U}(u_{AB})  & \tilde{I}(u_{AB})  - \tilde{V}(u_{AB})  
\end{pmatrix},
\end{equation}
where $u_{AB}$ is the coordinate of the Fourier plane where the baseline $A-B$ is observing. Plugging these equations in the definition of the trace (Eq. \ref{closure_trace}), we obtain the analytical expression of the closure trace for the interferometric observation of a double source presented in \citet{TFM}. 

\begin{figure}[!ht]
    \centering
    \resizebox{\hsize}{!}{\includegraphics{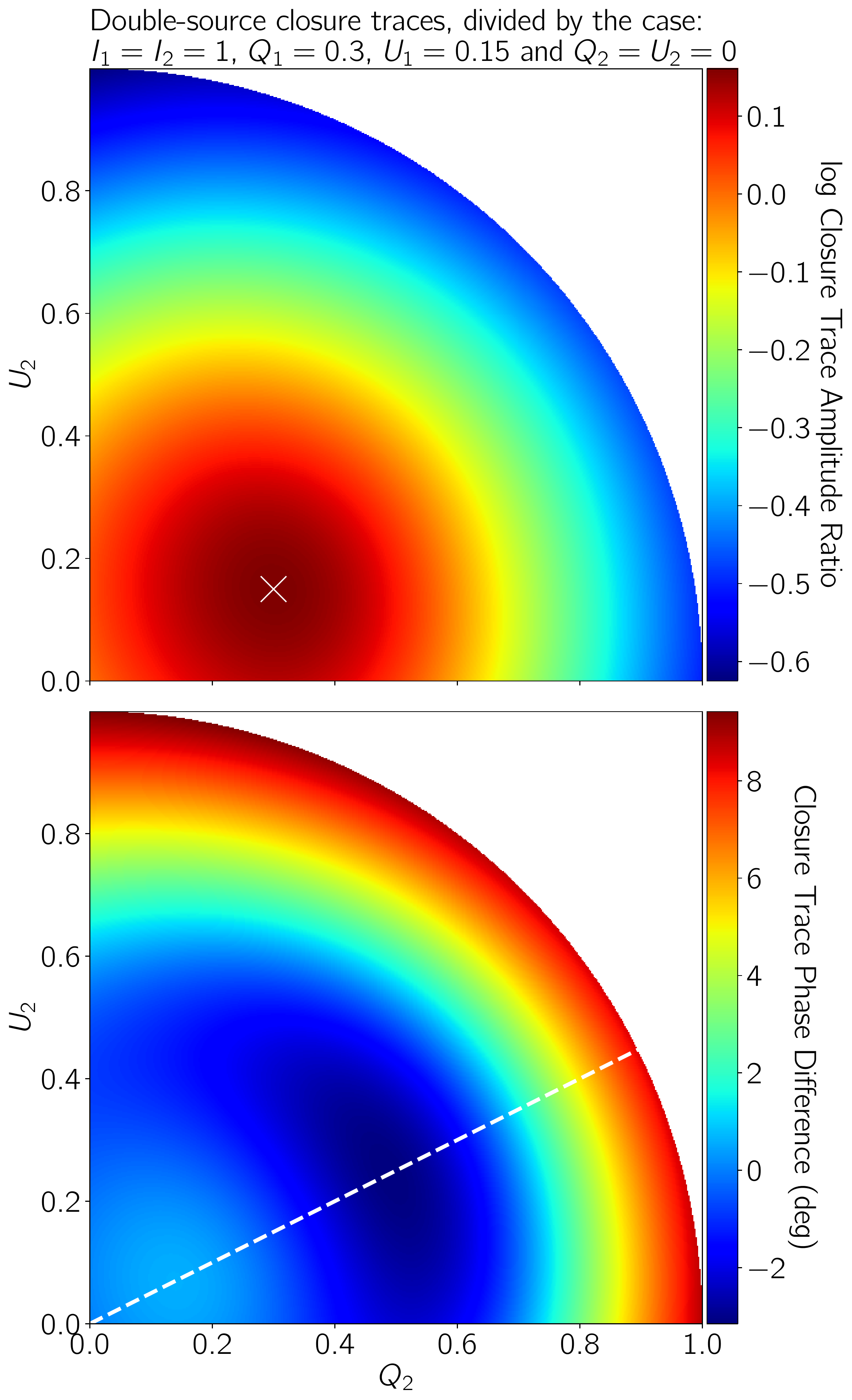}}
    \caption{Mapping of the closure trace phase difference and amplitude ratio of a double source, for different values of the Stokes parameters $Q_2$ and $U_2$, taking as reference trace the trace of the double source with $Q_2=U_2=0$ (i.e., unpolarized component $2$). The antennas are distributed in the positions $\{A;B;C;D\} \rightarrow \{0; 1; 4; 6\}$, in units of wavelength $\lambda$ for a source separation of $1.3$ radian. The white cross in the top panel marks the values of $Q_2$ and $U_2$ for which we get the maximum closure-trace amplitude ratio. On the other hand, the dashed line in the bottom panel marks the locus of points for which the electric vector position angle (EVPA) of component 1 and component 2 is aligned.}
    \label{fig:figA1}
\end{figure}

With a double source, we can analyse how the behaviour of the closure trace varies by modifying the Stokes parameters of one of the components, while keeping the polarization of the other constant. Hence, for a configuration of antennas at positions $\{A;B;C;D\} \rightarrow \{0; 1; 4; 6\}$, in units of wavelength $\lambda$, we fix the Stokes parameters of component $1$, $(I_1;Q_1;U_1; V_1) = (1;0.3;0.15;0)$, and we set $I_2 = I_1 = 1$ and $V_2 = V_1 = 0$. By doing this, we can give values to the Stokes parameters $(Q_2;U_2)$ and compute the closure traces, exploring the complete $(Q_2;U_2)$ parametric space. Finally, we compute the closure trace ratio, for each $(Q_2;U_2)$ values, relative to the closure trace of the double source with $Q_2=U_2=0$ (unpolarized component $2$). Thus, after generating a synthetic double source and computing the closure traces of a simulated observation, in Fig. \ref{fig:figA1} we show, for a certain quadruple of antennas, how both the closure traces phase and amplitude change as a function of Stokes $Q_2$ and $U_2$.

With this analytical exploration on the closure traces, we can highlight two conclusions. First, a closure trace amplitude peak (maximum or minimum) is obtained when the EVPAs of the two components are equal. Second, the closure trace phases are more sensitive to polarization intensity variations when the EVPAs of both components point in the same direction. In any case, another interesting conclusion is that the closure traces are actually sensitive to the differential polarization between the two components. Extending this conclusion to the more general case, if we observe frequency-dependent closure traces in an observation, it might be an indication of a frequency-dependent differential polarization (i.e., a spatially-resolved depolarization and/or RM). This paper continues to explore this idea, studying how closure traces could allow to extract information on the polarization structure of an observed source, and its frequency dependence directly from the observational data.

\end{appendix}

\end{document}